\documentclass[aps,prd,twocolumn,showpacs,showkeys,superscriptaddress,amsmath,amssymb]{revtex4}
\usepackage[usenames,dvipsnames]{color}
\usepackage{epsf}
\usepackage{bm}
\usepackage{dcolumn}
\usepackage{latexsym}
\usepackage{amsmath}
\usepackage{amsfonts}
\usepackage{amssymb}
\usepackage{graphicx}
\usepackage[active]{srcltx}

\newcommand{\be}{\begin{eqnarray}}
\newcommand{\ee}{\end{eqnarray}}

\def\d{\textrm{d}}
\def\bm#1{\mbox{\boldmath{$#1$}}}

\definecolor{darkred}{rgb}{.8,0,0}

\definecolor{darkblue}{rgb}{0,0,.7}

\begin{document}

\title{Tsallis thermostatics as a statistical physics of random chains}
\vspace{-2mm}

\author{Petr  Jizba}
 \email{p.jizba@fjfi.cvut.cz}
\affiliation{Faculty of Nuclear Sciences and Physical Engineering, Czech Technical University in Prague,
B\v{r}ehov\'{a} 7, 115 19 Praha 1, Czech Republic}
\affiliation{Institute for Theoretical Physics, Freie Universit\"{a}t in Berlin, Arnimallee 14, D-14195 Berlin, Germany}
\author{Jan Korbel}
\email{korbeja2@fjfi.cvut.cz}
\affiliation{Department of Physics, Zhejiang University, Hangzhou 310027, P. R. China}
\affiliation{Faculty of Nuclear Sciences and Physical Engineering, Czech Technical University in Prague,
B\v{r}ehov\'{a} 7, 115 19 Praha 1, Czech Republic}
\author{V\'{a}clav Zatloukal}
\email{vaclav.zatloukal@fjfi.cvut.cz}
\affiliation{Faculty of Nuclear Sciences and Physical Engineering, Czech Technical University in Prague,
B\v{r}ehov\'{a} 7, 115 19 Praha 1, Czech Republic}
\affiliation{Max Planck Institute for the History of Science, Boltzmannstrasse 22, 14195 Berlin, Germany}

\begin{abstract}
\vspace{3mm}
\begin{center}
{\bf Abstract}\\[2mm]
\end{center}
In this paper we point out that the generalized statistics of
Tsallis--Havrda--Charv\'{a}t  can be conveniently used as a conceptual
framework for statistical treatment of random chains.
In particular, we use the
path-integral approach to show that the ensuing partition function
can be identified with the partition function of a fluctuating oriented
random loop of arbitrary length and shape in a background scalar potential. To put some meat on the bare bones, we illustrate this with
two statistical systems; Schultz--Zimm polymer and relativistic particle.
Further salient issues such as the $PSL(2,{\mathbb{R}})$ transformation properties of
Tsallis' inverse-temperature parameter and a grand-canonical ensemble of fluctuating
random loops related to the Tsallis--Havrda--Charv\'{a}t statistics
%
are also briefly discussed.
\end{abstract}


\maketitle

\section{Introduction}

Over the past two decades, the Boltzmann--Gibbs (BG)
statistical mechanics has undergone an important conceptual shift. While
successful in describing stationary systems characterized by
ergodicity or metric transitivity, it fails to reproduce statistical behavior of many real-world systems in biology, astrophysics, geology, and the
economic and social sciences.  It is symptomatic of such
cases that one tries to find refuge in a new paradigm known as
``generalized statistics''.  The notion ``generalized statistics'' refers to statistical systems that are described with broad
(or fat-tail) distributions for which the usual Central Limit Theorem is inapplicable. Examples include generalized hyperbolic
distributions, Meixner distributions, Weibull, distributions, and various power-law tail distributions (e.g., Zipf--Pareto, L\'{e}vy,
Mandelbrot, or student-t (Tsallis-type) distributions).
The underlying mathematical underpinning and terminology involved are provided by various generalized Central Limit
Theorems (GCLTs), most notably by GCLT of P.~L\'{e}vy~\cite{levy,man,feller} and
B.V.~Gnedenko~\cite{gnedenko} or by non-Markovian GCLTs~\cite{[11c]}. In effect, GCLTs represent pertinent
frameworks incorporating such crucial theoretical concepts as L\'{e}vy processes, Mandelbrot's
L\'{e}vy flights~\cite{levy,man}, information-theoretic systems of R\'{e}nyi~\cite{Re2,PJ1}, non-extensive systems of
Tsallis--Havrda--Charv\'{a}t (THC)~\cite{HaCh,Ts2a} and their various generalizations~\cite{[11c],JKII,Thurner}.
Associated real-world phenomena obeying generalized statistics account for a rich class of statistical behaviors often observed in complex dynamical systems
ranging from financial markets~\cite{Mantegna,Borland,HK}, physics~\cite{[11c],Tsallis} and biology~\cite{Hansmann,HansmannII} to geoscience~\cite{Harte,HarteII}.

%
%
%
%
%

The aim of this paper is to point out that with the help the path integral (PI) one can
identify interesting new playgrounds for THC
generalized statistics.
In particular, the statistics in question is represented by distributions that emerge
when the maximal-entropy (MaxEnt) prescription is applied to
THC and R\'{e}nyi information measures (or entropies). Regardless
how narrow this class may seem, it constitutes an immense wealth of
real-world systems ranging from low-dimensional (at the edge of
chaos)~\cite{ECH1} and high-dimensional (self-organized
criticality)~\cite{SOC1} nonlinear dissipative systems, through well
developed turbulence~\cite{ArAr1} to long-range magnetic and
fluid-like systems~\cite{LRM1}. Here we wish to shed yet another light on
THC
generalized statistics by showing that it also represents a
pertinent framework for the statistical  physics of random chains.

Our paper is organized as follows: In Section~\ref{SEc2}
we present some essentials for both R\'{e}nyi and THC statistics
that will be needed in the main body of the paper. In Section~\ref{SEc2bc}, we
reveal and discuss the group structure of  the THC ``inverse  temperature'' parameter.
The group in question is the M\"{o}bius parabolic group which
is a one-parametric subgroup of the projective special linear group $PSL(2,\mathbb{R})$.
For the generalized statistics in question
we formulate the  relevant path-integral  representation for both the
density matrix and partition function. This is done
in Section~\ref{SEc3} with the help of Schwinger's trick.
In Section~\ref{SEc6} we show that the representation of the
density matrix obtained  naturally arises in the statistical theory of
random chains. We illustrate
our point with  the Schultz--Zimm polymer.
%
Important representatives of random chains are fluctuating particle histories
as encountered, for instance, in quantum mechanics or quantum field theory.
Remarkably, when we apply the generalized density matrix obtained to a free
{\em non-relativistic} particle in $D$ spatial dimensions we find that its is equivalent to
the (canonical Bloch) density matrix for a free {\em relativistic} particle in $D$ space-time dimensions
provided we set $q=2$ or $q=0$. This fact is proved in Section
\ref{SEc4} where also a generalization to a field-theoretical context is discussed.
%
Finally, Section~\ref{SEc7}  summarizes our results and discusses possible
extensions of the present work.
For the reader's convenience the paper is
supplemented with two appendices which clarify some finer technical details. In Appendix~A we derive
the Schulz--Zimm  distribution of chain lengths for relevant parameter values and in Appendix~B
we show how the generalized density matrix obtained can be seamlessly fitted into a computation
of the one-loop contribution to the Gibbs free energy in the case of  scalar quantum electrodynamics.

\section{Some fundamentals of R\'{e}nyi's and
THC statistics} \label{SEc2}

A useful conceptual frame that allows to generate
important classes of observed distributions is based on information
entropies. Information entropies generally represent measures of
uncertainty or ignorance inherent in a distribution describing a given
statistical or information-theoretical system. Central r\^{o}le of
information entropies is in that they serve as inference functionals
whose extremization subject to certain constraint conditions
(known as prior information) yields MaxEnt distributions
which often have direct phenomenological relevance.
Importance of information entropies as tools for inductive inference
(i.e., inference where new information is given in terms of expected
values) in statistical physics was emphasized by many authors
(see, e.g., references in~\cite{Fad1}), and presently MaxEnt approaches
belong among standard techniques from the statistical physics toolkit~\cite{[11c]}.

Among the many possible information entropies we shall
focus our attention here on two specific cases;
first on R\'{e}nyi's entropy~\cite{Re2} defined as
\begin{eqnarray}
{\mathcal{S}}_q^{(R)}  \ = \   \frac{1}{1-q} \ln \sum_i p_i^q\,
,\;\;\;\;\;\;\;\; q > 0\, ,\label{II1}
\end{eqnarray}
and second on the THC entropy~\cite{HaCh,Ts2a} which has the form
\begin{eqnarray}
{\mathcal{S}}_q^{(THC)}  \ = \  \frac{1}{1-q} \left(\sum_i p_i^q
-1\right)\, ,\;\;\;\;\;\;\;\; q > 0\, .\label{II2}
\end{eqnarray}
A discrete distribution ${\mathcal{P}} = \{p_i\}$ is usually
associated with a discrete set of microstates in statistical
physics, or set of all transmittable source symbols in information theory.
In the limit $q\rightarrow 1$, the two entropies coincide with each
other, both reducing to the Shannon (or Shannon--Gibbs) entropy
%
%
\begin{eqnarray}
S  \ = \  - \sum_{i} p_i \ln p_i\, .
\end{eqnarray}
In this way the parameter $q$ [or better $(q-1)$] characterizes the departure from the usual
BG statistics or from Shannon's information theory.

Let us remind that in the context of Shannon's
information theory it is well known~\cite{jaynes57} that the laws of equilibrium statistical mechanics can
be viewed as {\em inferences} based entirely on prior information
that is given in terms of expectation values of energy, energy and
number of particles, energy and volume, energy and angular momentum,
etc. In this case Shannon's entropy quantifies the information about the detailed microscopic state (microstate) of
the system, which remains ``uncommunicated'' by a description that is stated solely in terms of thermodynamic
state variables (phrased in terms of expectation values). It should be also stressed that the passage
from the Shannon--Gibbs to Clausius (i.e., thermodynamic) entropy
is established only when the relevant MaxEnt distribution
is inserted back into $S$. Only when this MaxEnt prescription is utilized then $S$ turn out to be a
thermodynamic {\em state function} and not a mere functional on a probability space.

In the spirit of MaxEnt strategy one can formally repeat the aforementioned philosophy
also for ${\mathcal{S}}_q^{(R)}$ and ${\mathcal{S}}_q^{(THC)}$.
It is still an open question, however, as to what extent the inferences obtained
can be identified with some genuine statistical system.
Here we do not wish to dispute a usefulness of ensuing MaxEnt distributions which clearly
serve as excellent fitting distributions in number of contexts in
complex dynamical systems~\cite{[11c],HK,Tsallis}. Instead, we wish to point out that
there are circumstances (which will be discussed shortly) when the ensuing generalized statistical
systems (and not only MaxEnt distributions) can be clearly identified with the real-world
equilibrium statistical systems.

For the sake of simplicity we shall proceed here only with the
analogue of canonical ensembles, where the prior information is
characterized by a fixed energy expectation value (i.e., the internal energy).
The corresponding MaxEnt distributions for ${\mathcal{S}}_q^{(R)}$ and
${\mathcal{S}}_q^{(THC)}$ can be obtained by extremization of the
associated inference function
\begin{eqnarray}
L_q^{(R;THC)}({\mathcal{P}}) &=& {\mathcal{S}}_q^{(R;THC)}  - \alpha
\sum_i p_i  -  \beta \langle H \rangle_r\, ,
\end{eqnarray}
where $\alpha$ and $\beta$ are Lagrange multipliers, the latter
being the analogue of the inverse temperature in natural units.
The subscript $r$ denotes the expectation value with respect to the weights
$P_i(r)  \equiv {p_i^r}/{\sum_j p_j^r}$ so that
\begin{equation}
\langle H \rangle_r \ = \ \sum_i P_i(r) E_i\, .
\label{5.aa}
\end{equation}
In practice it is common to chose only two values of $r$
that represent two different modes of application~\footnote{It also
reflects a certain confusion that still pervades the field. }.
In information theory one typically uses the linear mean
$\langle H \rangle_1 = \sum_i p_i E_i\,$ which corresponds to $r=1$,
while in non-extensive thermodynamics it is customary to utilize
a non-linear $q$-mean $\langle H \rangle_q$ which represents $r=q$. In the latter case is $P_i(q)$ called {\em escort\/}
or {\em zooming\/} distribution --- terminology that has its origin in chaotic dynamics~\cite{beck}.

In principle, we could keep the value of $r$ general when employing the MaxEnt procedure.
When this would be done, say for R\'{e}nyi's entropy, then
the MaxEnt distribution arises from the condition ${\delta L_q^{(R)}({\mathcal{P}})}/{\delta p_i}  =  0$. For generic $r$ this  leads to a higher-order trinomial equation for $p_k$, namely
\begin{eqnarray}
a {X_k^{(q-1)/(r-1)}} - b_k X_k -1 \ = \ 0\, ,
\label{6a}
\end{eqnarray}
when $r \neq 1$. In the special case when $r=1$ we have
\begin{eqnarray}
a X_k - b_k  -1 \ = \ 0\, .
\label{6b}
\end{eqnarray}
Here, $X_k = p_k^{r-1}$ in Eq.~(\ref{6a})  and $X_k =p_k^{q-1} $ in Eq.~(\ref{6b}). Parameters involved read
\begin{eqnarray}
 &&a \ = \ \frac{1}{Z_q} \ \equiv \ \frac{1}{\sum_l p_l^q}\, , \nonumber \\[1mm]
 &&b_k \ = \ \beta \ \! \frac{r\ \!(1-q)[E_k - \langle H \rangle_r ]}{qZ_r} \, .
\end{eqnarray}
Equations (\ref{6a}) are generally not suitable for physical considerations because they do not provide {\em unique} and {\em real} solutions.
There are only two cases in which the equations obtained are linear and thus yield a unique $p_k$, namely when $r =1$ and $r=q$.
We note in passing that for special values of $r$ and $q$ a solution of (\ref{6a}) of
the form
\begin{eqnarray}
p_k \ =  \ p_k(Z_r,Z_q, \beta,  \langle H \rangle_r; E_k)\, ,
\end{eqnarray}
can be found.
With two  constrains at hand we could eliminate $Z_q$ or $Z_r$ and rewrite $\beta$ in terms of
$\langle H \rangle_r$ (or, more physically,  $\langle H \rangle_r$ in terms of $\beta$).
The resulting $p_k$ would be, however, badly self-referential because either $Z_r$ or $Z_q$ would
not be eliminated. Again, cases $r =1$ and $r=q$  play a special r\^{o}le here since only in these
two cases
the vexing self-referentiality will be eliminated. This in turn provides a mathematical
(although not physical) backing for the two aforementioned choices of $r$ used in the literature.
With this proviso the associated MaxEnt distribution for R\'{e}nyi's entropy reads
\begin{eqnarray}
{p}_k^{R}  \ = \  Z_R^{-1}\left[ 1  - \beta^R_r (1+q-2r) \Delta_r E_k \right]^{{1}/{(1+q-2r)}}\, . \label{2.7}
\end{eqnarray}
Here $\Delta_r E_k = E_k - \langle H \rangle_r$, and $Z_R$ is the normalization constant
(basically the partition function). Term \mbox{$\beta^R_r = {{\beta r}/{q}}$} is the inverse
``temperature'' of the system.
%
%
By the same token one obtains for the THC case
%
\begin{eqnarray}
&&\mbox{\hspace{-4mm}}{p}_k^{THC} \! =  Z_{THC}^{-1}\left[ 1 \! - \beta^{THC}_r (1+q-2r) \Delta_r E_k \right]^{{1}/{(1+q-2r)}}\!,
\nonumber  \\
&&\mbox{\hspace{-4mm}}\label{2.8}
\end{eqnarray}
with the inverse ``temperature'' $\beta^{THC}_r  \! = \! \beta^R_r/\sum_i (p_i^{THC})^q$.
Similarly as before, this holds only when $r=1$ or $r=q$.
So, in contrast to (\ref{2.7}), the THC MaxEnt distributions are self-referential, i.e., $\beta^{THC}_r$ depends on the properties
of the distribution itself.
Distributions of the form (\ref{2.7}) and (\ref{2.8})
are known as $(2r-q)$-Gaussian or Tsallis' (thermostatistics) distributions
and they indeed appear in one form or another in numerous real-world statistical systems~\cite{Ts2a}.
For historical reasons the case $r=1$ in Eq.~\eqref{2.8} is also known as the Bashkirov's $1$-st version of
thermostatistics, while the case $r=q$ is called the Tsallis' $3$-rd version of thermostatistics~\cite{11}.

For future convenience  we note that the MaxEnt distributions
(\ref{2.7})--(\ref{2.8}) can be viewed as ${\mathfrak{q}}$-deformed versions of
the usual Gibbs--Boltzmann statistical distributions. For instance,
using the Box--Cox ${\mathfrak{q}}$-exponential~\cite{Box-Cox}, i.e.
\begin{equation}
e_{\{ {\mathfrak{q}}\}}^x  \ \equiv \  [ 1  +  (1- {\mathfrak{q}})x]^{1/(1-{\mathfrak{q}})}\, ,
\end{equation}
the resulting  $(2r-q)$-Gaussian distributions can be expressed in the succinct form
\begin{equation}
p(E_i) \ = \ Z^{-1} e_{\{2r-q\}}^{-\beta E_i}\, .
\end{equation}
Here $\beta$ represents corresponding inverse ``temperature'' and $Z$ is the normalization.
Since $e^x_{\{ 1\}} = e^x$, it is clear that Tsallis' thermostatistics distribution approaches in the limit
$q\rightarrow 2r-1$ the standard Maxwell--Boltzmann distribution of equilibrium statistical thermodynamics.

\section{More about $\beta^R_r$ and $\beta^{THC}_r $} \label{SEc2bc}

Though the distributions (\ref{2.7}) and (\ref{2.8}) have almost the same form, the main difference is
in the self-referentiality of $\beta^{THC}_r$. We will now illustrate that this subtle fact might have non-trivial (in fact measurable) consequences.
To this end we first recall the notable fact from the Boltzmann--Gibbs statistics, namely that the MaxEnt distribution is invariant under a constant energy shift $E_i \rightarrow E_i + \Delta$.
Actually, this result is not the consequence of a particular form of the BG entropy as one could think, but rather it directly results from a Legendre-transform structure and linear form of constraints~\cite{plas}.
With this proviso, the invariance under a constant energy shift is independent of the specific form of entropy and is also valid for $q$-escort constraints. This fact can be also directly observed in distributions (\ref{2.7}) and (\ref{2.8}) where the shift factor $\Delta$ does not appear in the spectrum-shifted distribution because the term $(E_i - \langle H \rangle_r)$ is manifestly $\Delta$ independent:
\begin{equation}
(E_i -  \langle H \rangle_r) \stackrel{+\Delta}{\rightarrow} [(E_i + \Delta) - \langle H + \Delta\rangle_r]  =  (E_i  - \langle H \rangle_r).
\end{equation}

Second, in the case of the BG distribution we can observe yet another important property, namely that it does not depend  on $\langle H \rangle \equiv \langle H \rangle_1$ {\em explicitly}. This follows from a simple factorization property:
\begin{equation}
\mbox{\hspace{-0.5mm}}p_i^{BG}  =  \frac{\exp[-\beta (E_i - \langle H \rangle)]}{\sum_j \exp[- \beta (E_j - \langle H \rangle)]} =  \frac{\exp(-\beta E_i)}{\sum_j \exp(- \beta E_j)}\, .
\end{equation}
%
We recall that the dependence on $\langle H \rangle$ is {\em implicitly} contained in $\beta$, because the distribution still has to fulfil the original  constraint~\footnote{In fact, in equilibrium thermodynamics the Lagrange multiplier $\beta$ turns out to be of key importance (even more important than the internal energy $\langle H \rangle$) so that physicist prefer to express $\langle H \rangle$ in terms of $\beta$ rather than $\beta$ in terms of $\langle H \rangle$. }
\begin{equation}\label{eq: constr}
\sum E_i\ \! p_i^{BG}(\beta, E_i) \ = \ \langle H \rangle\, ,
\end{equation}
which (in principle) allows to resolve $\beta$ in terms of $\langle H \rangle$. Let us remind that in the BG statistics
it is important to have the partition function without an explicit dependence on $\langle H \rangle$.
Explicit $\langle H \rangle$ dependence would, for instance, obscure the connection between the partition function and
Helmholtz free energy. In addition, by combining the constant-energy shift invariance of $p_i^{BG}$ with its lack of an explicit
$\langle H \rangle$ dependence one easily obtains the experimentally supported fact that the zero-point energy has no effect on the
BG statistical thermodynamics and can be thus safely set to zero.

It is a question, to what extent the {\em desired} absence of explicit $\langle H \rangle$  in $p_i^{BG}$ is inherited by distributions (\ref{2.7}) and (\ref{2.8}).
Contrary to the case with the constant-energy shift $E_i \rightarrow E_i + \Delta$, the factorization-like property for $\langle H \rangle_r$
is less trivial in Eqs.~(\ref{2.7})-(\ref{2.8}) and in fact it can be achieved only under special circumstances.


To see what is involved, let us assume that the second MaxEnt constraint (\ref{5.aa}) is not yet enforced, so that both the Lagrange multiplier $\beta$ and $\langle H \rangle_{r}$ are still independent. Within this framework any $\langle H \rangle_{r}$
appearing in considered distributions is explicit and the required lack of explicit dependence on $\langle H \rangle_{r}$
can be formulated as a form-invariance of the distribution under the shift of $\langle H \rangle_r$.
The MaxEnt constraint can be then safely imposed at the very end of our reasonings.

For $p_i^{BG}$
the form-invariance under the shift
\begin{eqnarray}
\langle H \rangle \rightarrow  \widetilde{\langle H \rangle}  = \langle H \rangle + {\mathfrak{b}}\, ,
\label{II.15.a}
\end{eqnarray}
(with ${\mathfrak{b}} \in \mathbb{R}$) is a simple consequence of the identity
\begin{eqnarray}
\mbox{\hspace{-0.5mm}}p_i^{BG}  &=&  \frac{\exp[-\beta (E_i - \langle H \rangle)]}{\sum_j \exp[- \beta (E_j - \langle H \rangle)]}\nonumber\\[1mm]
&=&  \frac{\exp[-\beta(E_i - \widetilde{\langle H \rangle})]}{\sum_j \exp[- \beta (E_j - \widetilde{\langle H \rangle})]}\, .
\end{eqnarray}
In this case $\beta$ is manifestly invariant under the shift (\ref{II.15.a}).

For THC MaxEnt distributions (\ref{2.8}) we might recall that $\beta^{THC}_r  = \beta_r^R/Z_{THC}(\beta, q,r, \langle H \rangle_r) = \beta^{THC}_r(\beta, q,r, \langle H \rangle_r)$. Since we shall ultimately be considering what happens under $\langle H \rangle_r$ translations, we can write only  $\beta^{THC}_r = \beta^{THC}_r(\langle H \rangle_r)$.
For further convenience we denote the shifted mean value $ \widetilde{\langle H \rangle}_r  = \langle H \rangle_r + {\mathfrak{b}}$ simply as ${\mathfrak{b}}$. We also set $\beta^{THC}_r = \beta$ and $Z_{THC} = Z$.

In the case of ${p}_k^{THC}$, the form-invariance under arbitrary shift in $\langle H \rangle_r$ can be achieved by compensating
for the shift by appropriately redefined/transformed form of $\beta^{THC}_r$.
Although this fact is known (cf. e.g., Ref.~\cite{tsallis:88}) it deserves further qualifications.
Our subsequent derivation will thus follow route less traveled but more suitable to our needs.

%
Aforementioned transformation properties of $\beta^{THC}_r$ can be read off from  the  identity
%
%
\begin{eqnarray}
&&\mbox{\hspace{-9mm}}\left[1  -  \beta^{}({\mathfrak{b}_1}) (1+q-2r)
(E_i + \mathfrak{b}_1)\right]\nonumber \\[2mm]
&&\mbox{\hspace{-9mm}}= \ \left[1 -
\beta^{}({\mathfrak{b}_2})(1+q-2r)(E_i +
\mathfrak{b}_2)\right] \ \! c({\mathfrak{b}_1},{\mathfrak{b}_2};q,r)\, ,
\label{SEc2.8a}
\end{eqnarray}
where $c({\mathfrak{b}_1},{\mathfrak{b}_2};q,r)$ is an energy-spectrum independent constant. From
Eq.~(\ref{SEc2.8a}) $c({\mathfrak{b}_1},{\mathfrak{b}_2};q,r)$ must fulfill two simultaneous equations
\begin{eqnarray}
&&\mbox{\hspace{-8mm}}c({\mathfrak{b}_1},{\mathfrak{b}_2};q,r) \ = \ {\beta^{}({\mathfrak{b}_1})}/{\beta^{}({\mathfrak{b}_2})}\, ,\nonumber \\
&&\mbox{\hspace{-8mm}}c({\mathfrak{b}_1},{\mathfrak{b}_2};q,r) \ = \ 1 - \beta^{}({\mathfrak{b}_1})(1+q-2r) (\mathfrak{b}_1-\mathfrak{b}_2)\, .
\label{2.18aa}
\end{eqnarray}
This implies that (\ref{SEc2.8a}) is satisfied only when ``inverse temperatures'' with different  ${\mathfrak{b}}$ obey the {\em linear fractional transformations}
%
\begin{eqnarray}
\beta({{\mathfrak{b}_2}})  &=&  \frac{\beta({{\mathfrak{b}_1}})}{1 -
\beta({{\mathfrak{b}_1}})(1+q-2r)(\mathfrak{b}_1-\mathfrak{b}_2)}\, ,\nonumber\\
\beta({{\mathfrak{b}_1}})  &=& \frac{\beta({{\mathfrak{b}_2}})}{1 -
\beta({{\mathfrak{b}_2}})(1+q-2r)(\mathfrak{b}_2-\mathfrak{b}_1)}\, .
\label{2.10}
\end{eqnarray}
%
Eq.~(\ref{SEc2.8a}) together with (\ref{2.10}) implies that the THC MaxEnt distribution is (form-)invariant under the shift of $\langle H \rangle_r$, namely
\begin{eqnarray}
&&\mbox{\hspace{-8mm}}\frac{\left(1-\beta^{}({{\mathfrak{b}_1}})(1+q-2r)(E_i +
\mathfrak{b}_1)\right)^{1/(1+q-2r)}}{Z(\beta^{}({{\mathfrak{b}_1}}))} \nonumber \\[1mm]
&&\mbox{\hspace{-6mm}}= \ \frac{\left(1-\beta^{}({{\mathfrak{b}_2}})(1+q-2r)(E_i +
\mathfrak{b}_2)\right)^{1/(1+q-2r)}}{Z(\beta^{}({{\mathfrak{b}_2}}))}\, .
\end{eqnarray}
%
%
Within this framework we can choose to work directly with $E_i$ rather than with $\Delta E_i$.

Let us remark that an analogous line of thoughts does not hold for R\'{e}nyi MaxEnt distributions (\ref{2.7}).
Indeed, consider, for instance the R\'{e}nyian analogue of Eq.~(\ref{2.10}) and set $\mathfrak{b}_1 = \langle H \rangle_r$.
In this case we have
\begin{eqnarray}
\beta_r^R  &=&  \frac{\beta_r^R({{\mathfrak{b}_2}})}{1 -
\beta_r^R({{\mathfrak{b}_2}})(1+q-2r)(\mathfrak{b}_2-\mathfrak{b}_1)}\, . \label{II.20.a}
\end{eqnarray}
By assuming that $\beta_r^R({{\mathfrak{b}_2}})$ is $\mathfrak{b}_1$ independent (otherwise ${p}_k^{R}$ would be $\langle H \rangle_r$-shift dependent) we can differentiate both sides of (\ref{II.20.a}) with respect to $\mathfrak{b}_1$, which yields identity
\begin{eqnarray}
0 \ = \ [\beta_r^R({{\mathfrak{b}_2}})]^2(1+q-2r)\, .
\end{eqnarray}
Since ${{\mathfrak{b}_2}}$ is by assumption arbitrary, this can be fulfilled only when $1+q=2r$. For $r$'s at hand the latter is equivalent to $q=1$, which in turn sends ${p}_k^{R}$ back to the BG distribution. The crux of course is, that $\beta_r^R$ (unlike $\beta_r^{THC}$) is not $\langle H \rangle_r$ dependent and thus it does not have means to compensate for the shift in $\langle H \rangle_r$.


%
Let us now turn back to ${p}_k^{THC} $ and observe that transformations~\eqref{2.10} constitute the
one-parameter subgroup of the projective special linear group $PSL(2,\mathbb{R})$.
This can be seen by realizing that
$PSL(2,\mathbb{R})$ is  the quotient group  $SL(2,\mathbb{R})/\mathbb{Z}_2$, where $\mathbb{Z}_2 = \{\pm {\bm 1} \}$ (${\bm 1}$ stands for $2\times 2$ unit matrix). The special linear group  $SL(2,\mathbb{R})$ is represented (in its fundamental representation) by $2\times2$ real matrices with
determinant one that act on a two-dimensional vector space as
\begin{eqnarray}
\left(
  \begin{array}{c}
    u_1 \\
    u_2 \\
  \end{array}\right) \; \mapsto \; \left(
                              \begin{array}{cc}
                                a & b \\
                                c & d \\
                              \end{array}
                            \right)\left(
                                     \begin{array}{c}
                                       u_1 \\
                                       u_2 \\
                                     \end{array}
                                   \right), \;\;\;\;\;\; ad  -  cb
 =  1\, .
\end{eqnarray}
If we now identify
\begin{eqnarray}
\left(
  \begin{array}{cc}
    a & b \\
    c & d \\
  \end{array}
\right)  = \pm \left(
                \begin{array}{cc}
                   1 & 0  \\
                  (2r -1-q)\mathfrak{b} & 1 \\
                \end{array}
              \right) \, \in \,
              SL(2,{\mathbb{R}})
              \, ,\label{13.bb}
\end{eqnarray}
and set $\beta({\mathfrak{b}}_i) = u_1({\mathfrak{b}}_i)/u_2({\mathfrak{b}}_i)$, we find that
$\beta({\mathfrak{b}}_i)$ transforms as~\eqref{2.10}. The parameter ${\mathfrak{b}}$ in (\ref{13.bb})
is the required value of the shift between initial and final configuration.

Alternatively, we might identify
\begin{eqnarray}
\left(
  \begin{array}{cc}
    a & b \\
    c & d \\
  \end{array}
\right)  =  \pm \left(
                \begin{array}{cc}
                  1 & (1+ q- 2r)\mathfrak{b}  \\
                   0 &  1 \\
                \end{array}
              \right) \, \in \,
              SL(2,{\mathbb{R}})
              \, ,\label{13.bc}
\end{eqnarray}
and set $\beta({\mathfrak{b}}_i) = u_2({\mathfrak{b}}_i)/u_1({\mathfrak{b}}_i)$. In this case we get the inverse transformation in terms of the original
parameter ${\mathfrak{b}}$. Both ensuing linear fractional transformations are, of course, equivalent which is on the level of (\ref{13.bb}) and (\ref{13.bc}) reflected by the fact that the respective matrices satisfy the equivalence relation (similarity transformation), mediated by the matrices
\begin{eqnarray}
&&\mathbb{S}_{++} \ = \ \mathbb{S}_{--} \ = \ \left(
                \begin{array}{cc}
                  0 & 1 \\
                   -1 & 0 \\
                \end{array}
              \right) ,\nonumber \\
              &&\mathbb{S}_{+-} \ = \ \mathbb{S}_{-+} \ = \ \left(
                \begin{array}{cc}
                  0 & i \\
                   -i & 0 \\
                \end{array}
              \right) .
\end{eqnarray}
For this reason it suffices to consider only one type of transformations, say (\ref{13.bb}).

From (\ref{13.bb}) and group of transformations (\ref{2.10}) we see that the relation is not one-to-one since any two matrices $\mathbb{A}$ and $-\mathbb{A}$ from (\ref{13.bb}) correspond to the same transformation in (\ref{2.10}).  Consequently, we have  a {\em homomorphism} from a subgroup (\ref{13.bb})  of $SL(2,\mathbb{R})$ (known as  $E(1)^{\pm}$) onto the group of  linear fractional transformations (\ref{2.10}) with kernel $\{\pm {\bm 1} \} = \mathbb{Z}_2$ and thus $E(1) = E(1)^{\pm}/\mathbb{Z}_2$ is isomorphic to group of transformations (\ref{2.10}). $E(1)$ is known as the {\em parabolic subgroup} of the projective group $PSL(2,\mathbb{R})$  or equivalently as the (real) {\em M\"{o}bius parabolic group}. A crucial point in this context is that
\begin{eqnarray}
\beta(0)  =  \frac{\beta({\mathfrak{b}})}{1  -
\beta({\mathfrak{b}})(1+q-2r){\mathfrak{b}}}\, ,  \;\;\;\;\;
\mbox{for}\;\; \;\; \forall {\mathfrak{b}}\in \mathbb{R}\, ,
\end{eqnarray}
represents an invariant under the above group of M\"{o}bius parabolic transformations --- \emph{Casimir invariant}.
In the following we denote the invariant quantity $[\beta(0)(2r-q-1)]^{-1}$ as $\mu$.
Such an invariant quantity represents a natural candidate for an {\em observable} or a {\em state variable}.
In fact, in number of cases it seems reasonable to identify $\mu$ directly with the statistical temperature. This step is, however, conceptually quite delicate because the thermodynamical temperature
(i.e., temperature obtained via Clausius or Caratheodory type of entropy-temperature duality) and
statistical temperature (i.e., temperature that appears as the Lagrange multiplier in MaxEnt distributions)
are a priori not related in R\'{e}nyi and THC statistics (unlike in Shannon-entropy based statistics).
In the context of the THC statistics above issue has been discussed by several authors~\cite{plas,Rama,Abe,Abe2,Martinez}. The strategy usually employed is to design (or engineer) the rules of thermodynamics (e.g., the first law of thermodynamics or Clausius theorem)
or Jaynes' MaxEnt reasoning so that both temperatures coincide.
In sections~\ref{SEc6} and \ref{SEc4} we will discuss the r\^{o}le of $\mu$ from a different angle of view. In particular,  we will see that is often more natural and conceptually less controversial to identify $\mu$ with other observables than temperature.

The $PSL(2,\mathbb{R})$ transformation properties of $\beta({\mathfrak{b}})$ are also non-trivially reflected in the scaling behavior of the partition function $Z(\beta({\mathfrak{b}}))$. In fact,  from (\ref{SEc2.8a})-(\ref{2.18aa}) we already know that
\begin{eqnarray}
&&\mbox{\hspace{-5mm}}\left[1-   \beta^{}({{\mathfrak{b}_2}})(1+q-2r) ({{\mathfrak{b}_1}} - {{\mathfrak{b}_2}}) \right]^{1/(2r -q -1)} Z(\beta^{}({{\mathfrak{b}_2}}))\nonumber \\[1mm]
&&\mbox{\hspace{-5mm}}=\ Z(\beta^{}({{\mathfrak{b}_1}}))\, .
\end{eqnarray}
Comparing this with (\ref{2.10}) we see that
\begin{eqnarray}
Z(\beta^{}({{\mathfrak{b}_1}})) \ = \ \left[\frac{\beta({{\mathfrak{b}_1}})}{\beta({{\mathfrak{b}_2}})}\right]^{1/(2r-q-1)} Z(\beta^{}({{\mathfrak{b}_2}}))\, .
\end{eqnarray}
This might be equivalently rewritten as
\begin{eqnarray}
Z(\lambda \beta^{}({{\mathfrak{b}}})) \ = \ \lambda^{1/(2r-q-1)} Z(\beta^{}({{\mathfrak{b}}}))\, ,
\end{eqnarray}
which shows the $Z(\beta^{}({{\mathfrak{b}}}))$ is a homogeneous function of degree $1/(2r-q-1)$ in $\beta^{}({{\mathfrak{b}}})$.
The solution can be clearly written in the form
\begin{eqnarray}
Z(\beta^{}({{\mathfrak{b}}})) \ = \ \beta^{}({{\mathfrak{b}}})^{1/(2r-q-1)} z(\mu)\, ,
\label{33abc}
\end{eqnarray}
where $z(\mu)$ is some function which depends on the invariant quantity $\mu$ (other state variables and coefficients  $q$ and $r$ are suppressed). This type of a power-like scaling law is characteristic for ensembles of fluctuating lines~\cite{HK,KleinertIII}. In this respect, path integrals provide the most natural tool for studying the statistical
fluctuations of line-like statistical systems~\cite{FH}.

\section{Path-integral representation of the THC density matrix}
\label{SEc3}

For our further reasonings it is convenient to consider
a quantum mechanical setting in which the probability
distribution is represented by the density operator. In such a case,
the (un-normalized) density operator associated with the THC
MaxEnt distribution reads
\begin{eqnarray}
\mbox{\hspace{-2mm}}\hat{\rho}(\beta({\mathfrak{b}}))  =  \left[ 1  -
\beta({\mathfrak{b}})(1+q-2r) (\hat{H} + {\mathfrak{b}})
\right]^{1/(1+q-2r)}\! \!\! . \label{3.1}
\end{eqnarray}

Let us now assume that $\hat{H}$ is the first-quantized Hamiltonian and that  $\beta({\mathfrak{b}})$ transforms according to $E(1)\subset PSL(2,\mathbb{R})$ group, so that one can compensate for the change in  ${\mathfrak{b}}$ by appropriately redefining $\beta({\mathfrak{b}})$.
For $q< 2r -1$  one can rewrite
$\hat{\rho}(\beta({\mathfrak{b}}))$ with a help of Schwinger's trick~\cite{HK,IZ} as
\begin{eqnarray}
\hat{\rho}(\beta({\mathfrak{b}})) =
\frac{1}{\Gamma(1/\varepsilon)} \int_{0}^{\infty} \frac{\d t}{t} \
t^{1/\varepsilon} e^{-t} e^{-\varepsilon \beta({\mathfrak{b}}) t
(\hat{H} + {\mathfrak{b}})}\, ,
\end{eqnarray}
with \mbox{$0 < \varepsilon = 2r - q-1$}. This allows to phrase the configuration-space density matrix
$\rho(x_a,x_b;\beta({\mathfrak{b}})) \equiv \langle
x_b|\hat{\rho}(\beta({\mathfrak{b}}))|x_a \rangle$ in the path-integral  form
\begin{widetext}
\begin{eqnarray}
\rho(x_a,x_b;\beta({\mathfrak{b}}))  =
\frac{1}{\Gamma(1/\varepsilon)} \int_{0}^{\infty} \frac{\d t}{t} \
t^{1/\varepsilon} e^{-t[1 + \varepsilon
\beta({\mathfrak{b}}){\mathfrak{b}}]} \int_{x(0) = x_a}^{x(\beta(t))
=x_b} {\mathcal{D}}x \int\! {\mathcal{D}}p \ e^{\int_{0}^{\beta(t)}
\d \tau (ip\dot{x} - H)}\, . \label{eq2.1}
\end{eqnarray}
\end{widetext}
Here $\beta(t) \equiv \varepsilon \beta({\mathfrak{b}}) t$.  In the case when
the potential-energy term in $H$ does not contain time derivatives of $x$, the momenta can
be integrated out leaving behind the usual
configuration-space path integral with the Euclidean action.
In particular
\begin{eqnarray}
\int_{0}^{\beta(t)}\!\!
\d \tau \! \ (ip\dot{x} - H)  \mapsto  - S_{e} = - \int_{0}^{\beta(t)}\!\!
\d \tau \! \ L_e (\dot{x},x)\, .
\end{eqnarray}
Here $S_{e}$ is the Euclidean action. For a Hamiltonian of the standard form $H(p,x) = p^2/2m + V(x)$ we would get the corresponding Euclidean  Lagrangian
in the form $L_e(\dot{x},x) = m \dot{x}^2/2 + V(x)$. Because of the plus sign in front of $V(x)$, the $L_e(\dot{x},x)$ is often denoted as
$H(\dot{x},x)$.

By changing variable $\beta(t) \mapsto \beta$ so that
$\beta = \varepsilon \beta({\mathfrak{b}})t$, we can
cast (\ref{eq2.1}) in the form
\begin{widetext}
\begin{eqnarray}
\rho(x_a,x_b;\beta({\mathfrak{b}}))  \ &=& \
\frac{1}{\Gamma(1/\varepsilon)\ \! [\varepsilon \beta({\mathfrak{b}})]^{1/\varepsilon}} \int_{0}^{\infty} \frac{\d
\beta}{\beta} \ \beta^{1/\varepsilon} e^{-\beta\mu} \int_{x(0) =
x_a}^{x(\beta) =x_b} {\mathcal{D}}x \int\! {\mathcal{D}}p \
e^{\int_{0}^{\beta} \d \tau
(ip\dot{x} - H)}\, ,\nonumber \\[3mm]
&=& \ \left[\frac{\beta({{0}})}{\beta({\mathfrak{b}})}\right]^{1/\varepsilon} \int_{0}^{\infty} \d \beta \ f_{\mu,
1/{\varepsilon}}(\beta)\int_{x(0) = x_a}^{x(\beta) =x_b}
{\mathcal{D}}x \int\! {\mathcal{D}}p \ e^{\int_{0}^{\beta} \d \tau
(ip\dot{x} - H)}\, .
 \label{3.2}
\end{eqnarray}
\end{widetext}
%
The smearing function
\begin{eqnarray}
\mbox{\hspace{-4mm}}f_{\alpha, \nu}(x)  =   \frac{1}{\Gamma(\nu)} \
\alpha^\nu x^{\nu -1} e^{-\alpha x}; \;\ \int_0^{\infty} \!\d x \
f_{\alpha, \nu}(x) = 1,
\label{gamma_PDF}
\end{eqnarray}
is the {\em Gamma} probability density function (PDF)~\cite{feller}. So, save for the multiplicative pre-factor,
the density matrix for the THC MaxEnt distribution can be viewed as the
Gibbsian density matrix weighted (or smeared) with the
Gamma-distribution. As expected, the ${\mathfrak{b}}$ dependence
entirely disappeared from the path integral expression in (\ref{3.2}) and it was replaced by the dependence on the invariant quantity $\mu$.

We note in passing that for large $\nu$ the following asymptotical behavior holds
\begin{eqnarray}
f_{\alpha, \nu}(x)  &\approx&  \sqrt{\frac{\alpha}{2\pi x}} \
 \left(\frac{\alpha x}{\nu}
 \right)^{\nu -1/2} e^{-\nu(\alpha x/\nu -1 )}\nonumber \\[2mm]
&\approx&
 \delta\!\left(x - \frac{\nu}{\alpha}\right)\,
 .
 \label{delta-func}
\end{eqnarray}
So for $q \rightarrow 2r -1$ the $\beta$-integration disappears and the  position-space density
matrix (\ref{3.2}) approaches the familiar PI representation of the (non-relativistic) Bloch density matrix  known from the BG statistics~\cite{FH}.

From (\ref{3.2}) follows that the corresponding partition
function can be written as
\begin{widetext}
\begin{eqnarray}
Z(\beta^{}({{\mathfrak{b}}})) \ = \  \int_{-\infty}^{\infty}\d x \ \rho(x,x;\beta({\mathfrak{b}})) \ = \  \left[\frac{\beta({{0}})}{\beta({\mathfrak{b}})}\right]^{1/\varepsilon}
 \int_{0}^{\infty} \d \beta \ f_{\mu,
1/{\varepsilon}}(\beta)\oint {\mathcal{D}}x \int\! {\mathcal{D}}p \
e^{\int_{0}^{\beta} \d \tau (ip\dot{x} - H)}\, ,\nonumber \\
%
%
\label{3.7a}
\end{eqnarray}
\end{widetext}
where the measure of integration is defined as
\begin{eqnarray}
\oint {\mathcal{D}}x \ \cdots  =  \int_{-\infty}^{\infty} \d x(0)
\int_{x(0) = x(\beta)} {\mathcal{D}}x \ \cdots \, .
\end{eqnarray}
Unfortunately, the partition function (\ref{3.7a}) overcounts number of physical
configurations. This is because translations $\tau \mapsto \tau +
const.$ do not change  the parametrization space which is now a
circle. This extra freedom does not allow to fix the starting point
${ x}$ on a loop uniquely and, in fact all choices are
equivalent. The rules of statistical physics prescribe that in the
partition function all equivalent configurations must be counted
only once if the theory is to make sense. Since for a loop of the
length $\beta$ we have $\beta$ different possibilities for a choice
of the starting point we must insert the extra factor $1/\beta$ in
${{Z}}$ to ensure that loops with different starting points $x(\tau)$
count as one loop.

So by defining
\begin{eqnarray}
&&\mbox{\hspace{-13mm}}z(\mu) \ \equiv \ \frac{1}{(\varepsilon \mu)^{1/\varepsilon}} \int_{0}^{\infty} \!\frac{\d
\beta}{\beta} \ f_{\mu, 1/{\varepsilon}}(\beta)\nonumber \\[1mm]
&&\mbox{\hspace{18mm}}\times \ \oint {\mathcal{D}}x\!\!
\int\!{\mathcal{D}}p \ e^{\int_{0}^{\beta} \d \tau (ip\dot{x} -
H)} \, ,
\label{43abc}
\end{eqnarray}
we may write the correct partition function as
\begin{eqnarray}
Z(\beta^{}({{\mathfrak{b}}})) \ = \ \beta^{}({{\mathfrak{b}}})^{-1/\varepsilon} z(\mu)\, . \label{3.7}
\end{eqnarray}
which we can recognize as the partition
function of a single fluctuating oriented random loop of arbitrary length and shape~\cite{KleinertIII,Parisi,Symanzik},
and correspondingly (\ref{3.2}) represents the density matrix of an open random chain with end-points $x_a$ and $x_b$ embedded in the loop.
In agreement with (\ref{43abc}), the loop lengths are distributed according to the Gamma PDF (\ref{gamma_PDF}) and
the chain interacts with a background scalar potential $V(x(\tau))$.
%
%
In this connection we should also remark that the integration parameter $\beta$ need no to be related to inverse temperature. Explicit examples of this fact will be illustrated in two subsequent sections. As anticipated, the form (\ref{3.7}) coincides with the formula (\ref{33abc}).

Frequently one is not interested in studying the behavior of a single fluctuating closed random chain but wants to consider grand-canonical ensembles of these. In such a case one can promote
the above THC statistics into a grand-canonical picture by exponentiating the single closed loop partition function (\ref{3.7}), so that the  grand-canonical partition function reads
\begin{eqnarray}
Z_{G} \ = \ e^{Z} \ = \ 1 + Z + \frac{1}{2!} Z^2 + \frac{1}{3!} Z^3 + \cdots \, .
\label{45aa}
\end{eqnarray}
This expansion comprises the no-loop, one-loop, two-loops, etc. contributions of mutually non-interacting
loops. A combinatorial factor $1/N!$ accounts for the indistinguishability of loops.
On account of (\ref{45aa}),  one may thus alternatively view $Z_{G}$ as the partition function of a {\em loop gas}.

Similarly as in the BG statistics, one should first multiply
the canonical partition function $Z$ with an arbitrary parameter ${\mathcal{M}}^{2(1/\varepsilon -1)}$ with
the dimension $2(1/\varepsilon -1)$ in mass units to make $Z$ dimensionless before it is inserted to (\ref{45aa}). (Here and throughout $\hbar = c = 1$).
%
%
With this proviso the partition function $Z_{G}$ can be written as
\begin{widetext}
\begin{eqnarray}
Z_{G}  &=&  \sum_{N=0}^{\infty} \frac{1}{N!} \prod_{k=1}^{N}\left[ \int_0^{\infty} \!\frac{\d
\beta_k}{\beta_k \Gamma(1/\varepsilon)} \ ({\mathcal{M}}^{2}\beta_k)^{1/\varepsilon - 1} e^{-\beta_k \mu}\oint {\mathcal{D}}x(\beta_k)\!\!
\int\!{\mathcal{D}}p(\beta_k) \right] \exp\left[\sum_{k=1}^N \int_{0}^{\beta_k} \d \tau_k \ \! (ip(\tau_k)\dot{x}(\tau_k) -
H(\tau_k)) \right]\nonumber \\[2mm]
&=&  \exp\left[ \int_0^{\infty} \!\frac{\d
\beta}{\beta \Gamma(1/\varepsilon)} \ ({\mathcal{M}}^{2}\beta)^{1/\varepsilon - 1} e^{-\beta \mu/{\mathcal{M}}^{2}}\ \! {\mbox{Tr}} \left(e^{-\beta \hat{H}}\right)\right]  =  \exp\left[ \frac{{\mathcal{M}}^{2s}}{s}\ \! \zeta_{[H +\mu]}(s)\right]  =  \exp\left[\frac{\zeta_{[(H + \mu)/{\mathcal{M}}^{2}]}(s)}{s}\right]\! ,
\label{46aa}
\end{eqnarray}
\end{widetext}
where we have set $s = (1/\varepsilon -1)$, defined the $\zeta_{[H + \mu]}$ function as
\begin{eqnarray}
\zeta_{[H + \mu]}(s) \ = \ \frac{1}{\Gamma(s)} \int_{0}^{\infty} \d\beta \ \! \beta^{s-1} \ \! {\mbox{Tr}} \left(e^{-\beta (\hat{H}+ \mu)}\right)\, ,
\end{eqnarray}
and used the scaling relation
\begin{eqnarray}
\zeta_{[(H + \mu)/{\mathcal{M}}^{2}]}(s) \ = \ {\mathcal{M}}^{2s}\zeta_{[(H + \mu)]}(s)\, .
\end{eqnarray}
%
%
Note also, that the multiplicative factors $\left[{\beta({{0}})}/{\beta({\mathfrak{b}})}\right]^{1/\varepsilon}$ and $\mu^{1/\varepsilon}$ were  
assimilated into the parameter ${\mathcal{M}}$.

With explicit representations (\ref{46aa}) at hand one can now employ various techniques and methodologies
used in the PI calculus to evaluate $Z_G$. This can be done either numerically (e.g., via PI Monte Carlo
or molecular dynamics simulations), in framework of approximative schemes (e.g., variational approaches or ergodic approximations~\cite{FH,Schulmann}) or via analytic perturbation schemes~\cite{HK,KleinertIII,JZI}. Apart from innate PI methods one can also employ operatorial approaches, such as Schwinger's perturbation expansion~\cite{Schwinger}  for ${\mbox{Tr}}\ \! e^{-\beta \hat{H}}$. We shall not dwell into these issues here, but instead we briefly mention another important perturbation treatment, namely the (spectral) $\zeta$-function expansion that is particularly pertinent in the framework of the THC statistics. The latter corresponds to an expansion of $Z_G$ around $\varepsilon = 1$, or equivalently around $q=0$ (for $r=1$) or $q=2$ (for $r=q$) and has a close connection with Quantum Field Theory (QFT). In particular, from the last identity in (\ref{46aa}) we can easily read out the expansion
\begin{eqnarray}
\mbox{\hspace{-0mm}}Z_G
 \! &=& \! e^{\left[\frac{1}{s} \zeta_{[(H + \mu)/{\mathcal{M}}^{2}]}(0) + \zeta_{[(H + \mu)/{\mathcal{M}}^{2}]}'(0) + \frac{s}{2} \zeta_{[(H + \mu)/{\mathcal{M}}^{2}]}''(0) + \ \! \ldots \right]}\nonumber\\[1mm]
&=& \!e^{\left[\zeta_{[H + \mu]}'(0) + \ln(M^2)\zeta_{[H + \mu]}(0) + \frac{s}{2} \zeta_{[(H + \mu)/{\mathcal{M}}^{2}]}''(0) \ \! + \ \! \ldots\right]}  \nonumber \\[1mm]
&=& \! {\det}\left[M^2(\hat{H} + \mu)^{-1}\right] e^{\left[\frac{s}{2e} \zeta_{[(H + \mu)/{M}^{2}]}''(0)\ \! + \ \! \ldots\right]} .
\label{48abc}
\end{eqnarray}
%
%
%
Here we have introduced
%
%
the dimensionful factor $M^2 = {\cal{M}}^2 \ \! e^{1/s}$
and employed the identity
\begin{eqnarray}
e^{\zeta_{[H + \mu]}'(0) + \ln(M^2)\zeta_{[H + \mu]}(0)}  \ = \ {\det}\left[M^2(\hat{H} + \mu)^{-1}\right] ,
\end{eqnarray}
[cf. Eq.~(\ref{Ap.B.3a}) in Appendix~B].  So, the leading contribution (in $s$) to $Z_G$ is easily recognized as the partition function of the complex scalar field theory~\cite{IZ}, namely
\begin{eqnarray}
&&\mbox{\hspace{-9mm}}{\mathcal{Z}} \ = \ {\det}\left[M^2(\hat{H} + \mu)^{-1}\right]  =   \int {\mathcal{D}}\phi^* {\mathcal{D}}\phi \ \! e^{-S[\phi^*,\phi ]} ,\label{50abc}\\[2mm]
&&\mbox{\hspace{-9mm}}S[\phi^*,\phi ] \ = \ \int  \d^D{\bm{x}} \ \! \phi^*({\bm{x}})(\hat{H} + \mu)\phi({\bm{x}})\nonumber \nonumber \\[1mm]
&&\mbox{\hspace{-5mm}} = \ \int  \d^D{\bm{x}} \ \! \phi^*({\bm{x}})\!\left(- \frac{1}{2m} \nabla^2 + \mu + V({\bm{x}})\right)\!\phi({\bm{x}})\, . \nonumber
\end{eqnarray}
The parameter $M$ is related in QFT to a normalization constant for ${\mathcal{Z}}$. 
In higher perturbation orders in (\ref{48abc}) it serves to absorb infinities arising from the behavior of $\beta$-integrals at small $\beta$. This short-distance  behavior 
(originally $[\beta] = [\mbox{kg}^{-2}] =  [\mbox{m}^{2}]$) can be
systematically dealt with via renormalization procedure.  

By using the vector notation ${\bm{x}}$ we emphasize the validity of our  reasoning also beyond $D=1$.  The field-theoretic partition functions (\ref{50abc}) typically appear
in the framework of effective field theories in which case
\begin{eqnarray}
\mbox{\hspace{-3mm}}V({\bm{x}}) = \left.\frac{\partial^2 U(\phi^*({\bm{x}}),\phi({\bm{x}}))}{\partial \phi^*{\bm{x}}
\partial \phi({\bm{x}})}\right|_{\phi({\bm{x}}) = \phi_c({\bm{x}}), \phi^*({\bm{x}}) = \phi^*_c({\bm{x}})}\! ,
\end{eqnarray}
where $U$ is the original field potential and $\phi_c({\bm{x}})$ together with $\phi^*_c({\bm{x}})$ are classical solutions
of the inceptive (i.e, non-effective) field theory.
Example of this type of behavior is illustrated in Appendix~B.

From aforesaid we see that the leading term in (\ref{48abc}) corresponds to the QFT representation of the loop gas in a background potential $V$. It should be noted that
because ${\d}(\beta^{s-1}/{\Gamma(s)} )/\d s = 1/\beta + {\mathcal{O}}(s)$, the loop lengths in the aforementioned  QFT representation
are distributed according to the exponential PDF (one-sided Laplace distribution) $e^{-\mu \beta} \mu$.
The sub-leading terms in $Z_G$ which are characterized by higher-order derivatives of $\zeta_{[(H + \mu)/{M}^{2}]}(s)$,
describe corrections to the exponential loop-length distribution in terms of powers of $s$ while 
keeping ${\mbox{Tr}}\ \! e^{-\beta \hat{H}}$ untouched. 
The latter is nothing but a variant of the Gram--Charlier expansion~\cite{Wal} of the THC-statistics-related Gamma PDF around the exponential distribution.


\section{Example~I: Schulz--Zimm model of polymer} \label{SEc6}

As mentioned, an important field of application of the foregoing formulas (\ref{3.2}) and
(\ref{3.7}) lies in the theory of random chains. Random chain of length $N$ is a sequence
$(\bm{x}_0,\ldots,\bm{x}_N)$ of $N+1$ points in a $D$-dimensional Euclidean space.
Each step $\Delta\bm{x}_n \equiv \bm{x}_n - \bm{x}_{n-1}$ ($n=1,\ldots,N$), i.e.,
bond connecting points $\bm{x}_{n-1}$ and $\bm{x}_n$, is a random variable with
fixed-length $|\Delta\bm{x}_n| = a$. Random chains are used to describe linear
molecular chains (polymers)~\cite{Edwards} as well as other line-like objects including vortex and defect lines
in condensed matter systems~\cite{KleinertIII,Polyakov}, fluctuating price histories in
financial markets~\cite{HK,JKH,KlKo} or fluctuating particle histories in quantum
mechanics~\cite{HK,Schulmann}.
In this section, we confine ourself to polymer chains.

Polymers are chemical compounds consisting of a large number of monomer units that are linked together by chemical bonds.
Examples include DNA, proteins, cellulose, sugars, rubber, etc.
There is a natural framework for modeling polymers in probability theory and statistical
physics~\cite{HK,KleinertIII,Edwards}:
a polymer chain is modeled by a random path with a probability distribution that is
Gibbsian; more specifically, one defines an energy functional on polymer configurations such that the higher
is the energy of the configuration, the less likely it appears.

Motivated by different physical phenomena, a variety of polymer models have been proposed
and studied in the probability and statistical physics literature.
A large class of polymers behaves approximately as \emph{ideal random chains} in which
case the links $\Delta\bm{x}_n$ are independent, identically distributed random variables,
taking values uniformly over a sphere of radius $a$. The parameter $a$ is known as the bond length of the random chain.
The probability distribution of the
end-to-end vector $\bm{R} \equiv \bm{x}_N - \bm{x}_0$ for an ideal chain of length $N$ can be thus written as~\cite{HK}
%
\begin{eqnarray}
&&\mbox{\hspace{-10mm}}P_N(\bm{R}) \ = \ \prod_{n=1}^N \left[ \int \d \Delta \bm{x}_n \frac{1}{S_D a^{D-1}} \! \
\delta\left(|\Delta\bm{x}_n|-a\right) \right]\nonumber \\[1mm]
&&\mbox{\hspace{2mm}}\times \ \delta^{(D)}(\bm{R} - \sum_{n=1}^N \Delta\bm{x}_n)\, ,
\label{A.32.a}
\end{eqnarray}
%
with $S_D = {2 \pi^{D/2}}/{\Gamma(D/2)}$ being the surface of a unit sphere in $D$ dimensional
space. In the limit of large $N$, $P_N(\bm{R})$ can be approximated, as a consequence of the
central limit theorem~\cite{feller,gnedenko,[11c]} by the Gaussian distribution
\begin{equation}
P_N(\bm{R}) \ \approx \ \left(\frac{D}{2 \pi a L}\right)^\frac{D}{2} \exp\left( -\frac{D \bm{R}^2}{2 a L} \right) \ \equiv \ P_L(\bm{R}) \, .
\label{end-to-end distribution ideal}
\end{equation}
In the following we shall use the actual polymer length $L=N a$ instead of diverging $N$.
Relation~(\ref{end-to-end distribution ideal}) may be understood as the propagator of a free
non-relativistic particle of a mass $m =D/a$ with time continued to an imaginary value $-iL$, i.e.
\begin{equation}
P_L(\bm{R}) \ = \ \langle\bm{x}_N, t_N |\bm{x}_0, t_0  \rangle|_{t_N - t_0 = -iL}\, .
\end{equation}
The corresponding path-integral representation of (\ref{end-to-end distribution ideal}) is in
the polymer literature known as Edwards integral~\cite{Edwards} and reads
\begin{equation} \label{ideal chain end-to-end distribution}
\mbox{\hspace{-2mm}}P_L(\bm{R})  =  \int_{\bm{x}(0)=\bm{0}}^{\bm{x}(L)=\bm{R}} \mathcal{D}\bm{x}
\exp \left( -\int_0^L H(\bm{\dot{x}}(\tau)) \d\tau \right)\! ,
\end{equation}
where
\begin{equation} \label{ideal chain hamiltonian}
H(\bm{\dot{x}}) \ = \ \frac{D}{2 a} \ \! \bm{\dot{x}}^2\, ,
\end{equation}
plays the r\^{o}le of the energy density of the polymer conformation $\bm{x}(\tau)$.

In real polymers, the bonds usually do not allow for an equal probability of all spherical
angles because the chains are stiff. To account for stiffness, one may use the coarse-graining
trick and increase, for sufficiently long chains, the bond length $a$ in
(\ref{ideal chain hamiltonian}) to the effective bond length $a_{\textrm{eff}}$, so
that chain segments of length $a_{\textrm{eff}}$ behave as freely rotating.

So far we have considered polymers of a fixed length $L$ (or a fixed degree of polymerization $N$).
In real solutions, however, various chain lengths are found with their distribution depending on
the nature of the polymerization reaction. For linear addition polymerization with termination,
where identical monomers are added one at a time to the reactive end of a growing chain until
the process is terminated, the \emph{Schulz--Zimm} (or \emph{Schulz})
PDF is typically used~\cite{Schulz,Zimm,Elias} (for further details see Appendix~A). It has a form of the Gamma-distribution (\ref{gamma_PDF})
with ${\nu}/{\alpha} = \langle L \rangle \equiv \bar{L}_n$ representing averaged polymer
length (known as the number-average molecular weight) and
\begin{eqnarray}
\frac{\nu+1}{\alpha} \ = \ \frac{\langle L^2 \rangle}{\langle L \rangle} \ \equiv \ \bar{L}_w\, ,
\end{eqnarray}
denoting the so-called weight-average molecular weight. The relative fluctuation (or better variance) reads
\begin{eqnarray}
\frac{\mbox{Var}(L)}{\langle L \rangle} \ = \ \frac{\bar{L}_w}{\bar{L}_n} - 1 \ = \ \frac{1}{\nu}\, .
\end{eqnarray}
The fraction ${\bar{L}_w}/{\bar{L}_n}$ is known as a polydispersity index (PDI) and it quantifies
the width of the distribution of molecular lengths. It equals $1$ for uniform chain length and is greater
than $1$ otherwise. The Schulz--Zimm PDF is used primarily because it has a simple functional form that spans the
monodisperse distribution (PDI~$ = 1$) and the most probable distribution (PDI~$ = 2$).

If the molecular chain can have any length with a distribution $\omega(L)$, the
length distribution of the end-to-end vector of ideal polymer is then
%
\begin{eqnarray}
&&\mbox{\hspace{-10mm}}P(\bm{R}) \ = \ \int_0^\infty \d L \ \! \omega(L) \nonumber \\[1mm]
&&\mbox{\hspace{0mm}}\times \ \int_{\bm{x}(0)=\bm{0}}^{\bm{x}(L)=
\bm{R}} \mathcal{D} \bm{x} \exp \left( -\int_0^L H(\bm{\dot{x}}(\tau)) \d\tau \right)\! .
\label{end-to-end vector distribution}
\end{eqnarray}
%
In the case of the Schulz--Zimm length-smearing distribution $f_{\alpha,\nu}(L)$ the above
$P(\bm{R}) $ coincides with the Tsallis' density matrix (\ref{3.2}) of a quantum particle described by Hamiltonian $H$.
The r\^{o}le of the smearing parameter $\beta$ is then played by the chain length $L$, the parameter
$\nu = {1}/{\varepsilon}$, and $\alpha = \mu = [{\varepsilon\beta(0)}]^{-1}$. We identify the
Tsallis' $q$-parameter with the polydispersity index of the distribution of polymer lengths
${\bar{L}_w}/{\bar{L}_n}$, and the Tsallis' temperature $\beta(0)$ with the mean chain length $\bar{L}_n$.

Let us finally add few comments. By modifying the law of the random walk (\ref{A.32.a})
(or alternatively (\ref{ideal chain end-to-end distribution})) via the introduction of an appropriate
energy functional, more realistic features can be introduced to account for the interaction between
different monomers, and the interaction between the polymer and the environment. Furthermore,
randomness (i.e., disorder) can be incorporated into such interactions to model impurities. This
leads to challenging problems, rich both in physical behavior and in  mathematical structure.

Let us also note that the r\^{o}le of Tsallis'  $E(1) \subset PSL(2,\mathbb{R})$ invariant
$\mu$ is played, according to  Appendix~A,
by $w_T/a$ where $w_T$ is the probability that a new monomer cannot be added to
a polymer chain (the chain is inactive).
However, in several dimensions it is also
possible to formulate the famous excluded-volume problem, which is the random walk
with the additional stipulation that no lattice point can be occupied more
than once (self-avoiding walk). This model is used as a simplified description
of a polymer: each monomer can have any position in space, given only the
fixed length of the links and the fact that no two monomers can overlap.
This problem has been the subject of extensive approximate, numerical, and asymptotic studies.
A fully satisfactory solution of the problem, however, has not been found. The
difficulty is that the model is essentially non-Markovian: the probability
distribution of the position of the next monomer depends not only on
the previous one or two, but on all previous positions.



\section{Example~II: connection with relativistic particles} \label{SEc4}

In a sense the simplest representative of the THC statistics-based random chains are fluctuating relativistic particle orbits.
In particular, the density matrix (\ref{3.2}) together with the partition function (\ref{3.7})
can be identified with the density matrix and partition function for a free spinless relativistic particle, respectively,
provided we use a suitable $S_{e}$ and set $q = 2r-2$.

The simplest place to start is the Polyakov-type
action for a free spinless particle which reads~\cite{HK,Teitelboim,PJ-HK,PJ-FS}
\begin{eqnarray}
&&\mbox{\hspace{-10mm}}S[x, \eta; \tau_1,\tau_2]\nonumber \\[2mm]
&&\mbox{\hspace{-10mm}}=  -
\mbox{$\frac{1}{2}$}\int_{\tau_1}^{\tau_2} \d\tau \
\left(\eta^{-1}(\tau)\ \!\dot{x}^{\mu}(\tau)\dot{x}_{\mu}(\tau) +
\eta(\tau)\ \! m_0^2 \right)\, . \label{3.8}
\end{eqnarray}
Here $\eta$ represents the square root of the world-line metric
(i.e., einbein) and $\tau$ is a label time (be it proper time,
affine parameter, etc.) parametrizing the world-line. We have chosen
the Lorentz signature in $D$ dimensions to be $(+, -, -, \cdots,
-)$. The action (\ref{3.8}) is invariant under reparametrizations of
the label time, i.e.,
\begin{eqnarray}
&&\tau \ \mapsto \ \bar{\tau}= f(\tau)  \;\; \Rightarrow \;\;
\dot{x}_{\mu}\dot{x}^{\mu}  \mapsto
\dot{\bar{x}}_{\mu}\dot{\bar{x}}^{\mu}  =
\frac{\dot{x}_{\mu}\dot{x}^{\mu}}{\dot{f}^2}; \nonumber \\
&&\d\tau \ \mapsto \ \d\bar{\tau}  =  \d\tau \dot{f}; \;\;\; \eta
\mapsto \bar{\eta} = \frac{\eta}{\dot{f}}\, , \label{3.9}
\end{eqnarray}
with $f(\tau)$ fulfilling the conditions $f(\tau_1)= \tau_1$,
$f(\tau_2)= \tau_2$ and $\dot{f} > 0$. In result  $S = \bar{S}$. The
transition amplitude from $x^{\mu}(\tau_1) = x^{\mu}_1$ to
$x^{\mu}(\tau_2) = x^{\mu}_2$ can be then written as
\begin{eqnarray}
\langle x_2, \tau_2| x_1,\tau_1 \rangle  =  \int_{x(\tau_1) =
x_1}^{x(\tau_2) = x_2} {\mathcal{D}} x \int {\mathcal{D}}\eta \
e^{iS}\, .\label{3.10}
\end{eqnarray}
This path integral is however not quite right. It contains an
enormous overcounting, because configurations $(\eta, x)$ and
$(\bar{\eta}, \bar{x})$ that are related to one another by the
reparametrization transformation (\ref{3.9}) represent the same
physical configuration. If we define the space of all einbeins
$\eta$  as $\Sigma$ and the $D$ dimensional Minkowski space as
${\mathbb{R}}_{\mbox{\tiny{M}}}^{\mbox{\tiny{D}}}$ then the true
space of physical configurations is not $\Sigma \times
{\mathbb{R}}_{\mbox{\tiny{M}}}^{\mbox{\tiny{D}}}$ but rather the factor
space $(\Sigma \times
{\mathbb{R}}_{\mbox{\tiny{M}}}^{\mbox{\tiny{D}}})/G$ with $G$
representing the reparametrization group. At least locally we can
always write that $\Sigma \times
{\mathbb{R}}_{\mbox{\tiny{M}}}^{\mbox{\tiny{D}}} \sim \left((\Sigma
\times
{\mathbb{R}}_{\mbox{\tiny{M}}}^{{\mbox{\tiny{D}}}})/G\right)\!\times
G$. Thus
\begin{eqnarray}
{\mathcal{D}}\eta {\mathcal{D}} x  =  \d \mu\!\left((\Sigma \times
{\mathbb{R}}_{\mbox{\tiny{M}}}^{{\mbox{\tiny{D}}}})/G\right) \ \d
\mu(G)\, .
\end{eqnarray}
Here $\d \mu(G)$ represents the measure on the reparametrization
group. It can be shown~\cite{Polyakov} that
\begin{eqnarray}
{\mathcal{D}}\eta {\mathcal{D}} x  &=&  \frac{\d L
{\mathcal{D}}x}{\sqrt{L}} \ \sqrt{{\det}_R\left( - \frac{\d^2
}{\d\tau^2} \right)}\ \d \mu(G)\nonumber \\[2mm]
&=& {\mathcal{N}}\ \!{\d L {\mathcal{D}}x}\ \d \mu(G) \, .
\end{eqnarray}
(Subscript $R$ means regularized determinant.) The variable $L$
corresponds to the actual length of the world-line, i.e.,
\begin{eqnarray}
L  =  \int_{\tau_1}^{\tau_2} \d \tau \ \!\eta(\tau) \, .
\end{eqnarray}
After factorizing out the volume of the reparametrization group we
obtain the true transition amplitude (\ref{3.10}) which now reads
\begin{eqnarray}
\langle x_2, \tau_2| x_1, \tau_1 \rangle  &=&  \int d
\mu\!\left((\Sigma \times
{\mathbb{R}}_{\mbox{\tiny{M}}}^{{\mbox{\tiny{D}}}})/G\right) \
e^{iS}\nonumber \\
&=&  {\mathcal{N}}\int_{0}^{\infty} \!\d L \ e^{-i m_0^2 L/2}
\int_{x(\tau_1) =
x_1}^{x(\tau_2) = x_2} {\mathcal{D}}x   \ e^{i \tilde{S}}\, , \nonumber \\[1mm]
&&\mbox{\hspace{-27mm}}\tilde{S}[x;\tau_1, \tau_2]  =  -
\mbox{$\frac{1}{2}$}\int_{\tau_1}^{\tau_2} \!\d \tau \
\dot{x}^{\mu}(\tau)\dot{x}_{\mu}(\tau)\, .\label{3.11}
\end{eqnarray}
Eq.(\ref{3.11}) is the so called world-line representation
of Green's function for Klein--Gordon equation and formally it might
be obtained also via the Feynman--Fock fifth parameter
approach~\cite{Schulmann}. Result (\ref{3.11}) can be naturally
related to the  density matrix by the substitution $\tau \mapsto -i
{\mathfrak{t}}$ and $L  \mapsto -i \beta $. In such a case we arrive
at the density matrix
\begin{eqnarray}
&&\rho(x_a,x_b;\beta)  =  {\mathcal{N}}\int_{0}^{\infty} \!\d
\beta \ e^{-
m_0^2 \beta/2} \int_{x(0) = x_1}^{x(\beta) = x_2} {\mathcal{D}}x    \ e^{- \tilde{S}_e}\, ,\nonumber \\[1mm]
&&\tilde{S}_e[x,\beta]  =  \mbox{$\frac{1}{2}$}\int_{0}^{\beta}
\!\d {\mathfrak{t}}
\left(\dot{\bf{x}}({\mathfrak{t}})\cdot\dot{\bf{x}}({\mathfrak{t}})
+ \dot{{x}}_0({\mathfrak{t}})\dot{x}_0({\mathfrak{t}})\right)\,
, \label{3.12}
\end{eqnarray}
with $\tilde{S}_e$ representing the ensuing Euclidean action.

Comparing (\ref{3.12}) with (\ref{eq2.1}) we see that Tsallis'
density matrix for a free {\em non-relativistic} particle in $D$
spatial dimensions is equivalent to the (canonical) density matrix
for a free {\em relativistic} particle in $D$ space-time dimensions
provided we identify $2r-q=2$ and $m_0^2/2 = \mu$. Because $PSL(2,{\mathbb{R}})
\simeq SO^+(1,2) \subset SO^+(1,D-1)$
(here $SO^+(1,D-1)$ represents the restricted Lorentz group in $D$ dimensions)
we have that $\mu$ is a Lorentz invariant in $D$ dimensions
(since $m_0^2$ is) and hence it is automatically invariant also under the
subgroup $PSL(2,{\mathbb{R}})$. This ensures
that (\ref{3.12}) agrees with the THC density matrix.

With the density matrix at hand we
can construct the corresponding one-particle partition function
${{Z}}$. Once again we have to be careful and insert the extra factor $1/\beta$
in the path-integral measure to avoid overcounting loops with different starting
points $\bm{x}({\mathfrak{t}}) = (x_0({\mathfrak{t}}), {\bf x}({\mathfrak{t}}))$.

The THC density matrix (\ref{3.12}) was constructed on the premise that $S$
(and ensuing $\tilde{S}_e$) describes a single particle. Of course, the single-particle relativistic
quantum theory is logically untenable, since a
multi-particle production is allowed whenever a particle
reaches the threshold energy for pair production.
So, strictly speaking the representation (\ref{3.12}) holds only when
energy-momentum involved is lower than the particle's rest mass.
In addition, Leutwyler's no-interaction theorem~\cite{Leut} prohibits
interaction for any finite number of particles in the context
of relativistic mechanics. To get around the no-interaction theorem
it is essential to have an infinite number of degrees
of freedom to describe interaction. The latter is typically
achieved via local quantum field theories (QFTs).

Despite aforesaid shortcomings, it should be stressed that the path integrals for a single
relativistic particle (\ref{3.11})-(\ref{3.12}) represent a key building block
in QFT. In fact, QFT in general, can be viewed
as a grand-canonical ensemble of fluctuating particle histories
(world-lines) where Feynman diagrammatic representation
of quantum fields depicts directly the pictures of
the world-lines in a grand-canonical ensemble. In particular,
the partition function for quantized relativistic fields
can be fully rephrased in terms of single-particle relativistic
PIs. This, the so-called ``world-line quantization'' of particle
physics, is epitomized, e.g., in Feynman's worldline representation
of the one-loop effective action in quantum electrodynamics~\cite{Feynman},
in the Strassler and Bern--Kosower  ``string-inspired'' approaches
to QFT~\cite{Strassler,Bern} or in disorder field theory~\cite{KleinertIII}.


Let us finally mention that the similar analysis we have
just done for a spinless relativistic particle could be
straightforwardly generalized to any spinning relativistic particle
with a non-zero rest mass (massive Rarita--Schwinger particle)~\cite{PJ-HK}. It
is also trivial to extend our approach to account for charged
relativistic particles that are coupled the external electromagnetic
field~\cite{PJ-HK}. This can be done as usual by the minimal substitution (via
covariant derivative). To put some meat on the bare bones, we use in Appendix~B the
world-line quantization to calculate   the Gibbs free energy, of the charged scalar quantum field in background electromagnetic
potential. At present, it seems that massless particles
do not fit easily in the outlined THC statistical scenario.


\section{Conclusions} \label{SEc7}
%

In this article we have introduced a class of
stochastic processes in which Tsallis' thermostatistics finds its natural conceptual playground.
The processes in question are characterized by the position-space  density matrix
which is invariant under the group $E(1) \subset PSL(2,\mathbb{R})$  (i.e., group of M\"{o}bius parabolic transformations) of the  THC ``inverse-temperature'' parameter.
We have seen that such a behavior is dictated by purely thermodynamic considerations (the first law of thermodynamics alongside with Legendre structure)
and supported by a self-referentiality of the underlying THC distribution (\ref{2.8}).
In contrast with the THC MaxEnt distribution, the closely related R\'{e}nyi MaxEnt distribution (\ref{2.7}), though with superficially identical form, is not
self-referential and consequently cannot compensate for the shift  of $\langle H \rangle_r$ by redefining consistently $\beta$.

We have shown that statistical systems that fit the above pattern of behavior can be
identified with certain types of random chains in a background scalar potential. This could be seen particularly clearly when
the associated density matrices are formulated in the path-integral language. In such a case
the ensuing partition function coincides with the partition function of a fluctuating
random loop of arbitrary length, while the density matrix itself describes an open random chain with end-points embedded in the loop.
A specific point of the THC statistics (namely its $E(1)$ symmetry) is that the loop lengths are distributed according to the Gamma PDF.
As an illustration of the issues involved we presented a treatment of two simple statistical systems, namely
an ensemble of fluctuating polymer chains in a Schulz--Zimm approximation
and ensemble of relativistic particle orbits formulated in the framework of relativistic quantum mechanics.

The PI representation of the THC density matrix also serves as a convenient starting point for various generalizations.
In particular, it provides a natural passage
from a single THC statistics-based fluctuating random loop to grand-canonical ensembles of these. We have discussed the basic inner workings of this procedure and highlighted its connection 
with the spectral $\zeta$-function expansion and ensuing QFT representation of the loop gas.
A closely related computation of the Gibbs free energy in scalar quantum electrodynamics
was presented in Appendix~B. Particularly in this latter case we could observe yet another r\^{o}le of the THC parameter, namely  it could be identified with a regulator in the $\zeta$-function regularization of functional determinants.

\begin{acknowledgments}
It is pleasure to acknowledge helpful conversations with
T.~Arimitsu, H.~Kleinert and D.~\v{S}tys.
PJ and JK were supported by the GA\v{C}R Grant No. GA14-07983S
and VZ by the DFG Grant KL 256/54-1.
\end{acknowledgments}

 \section*{Appendix A~\label{ap1}}
%

%

Here we provide a brief derivation of the Schulz--Zimm  distribution of chain lengths
for the values of the parameter $\nu = 1, 2$. We shall adopt an effective picture of
Kamide and Dobashi~\cite{Kamide} in which there is a certain probability $w_P$
of adding a monomer to the growing chain, and a probability $w_T = 1-w_P$ that
the termination reaction occurs such that monomers cannot be added anymore, i.e., the chain is dead.
The reader may find a detailed description of the chemical processes involved, e.g., in Ref.~\cite{Hiemenz}.
A chain composed of $n$ monomers ($n \geq 1$) is thus formed with the probability
$p(n) = w_P^{n-1} w_T$. There are two common processes of termination:
\emph{disproportionation} when two growing chains meet to deactivate their
reactive centers, but don't combine; and \emph{combination} when two growing chains
of lengths $n$ and $m$ combine to form a dead chain of length $n+m$.

In the case of termination by disproportionation, the distribution of chain lengths
is given simply by $p_D(n) = p(n) =  w_P^{n-1} w_T$. In the continuum limit, $L = n a$
($a$ is the bond length) and we use the approximation $w_P \approx 1$, i.e. $w_T \ll 1$,
in which long chains are likely to be formed. The probability density $P_D(L)$ to find a chain of
length $L$ is determined from the identity
\begin{eqnarray}
p_D(n) \ = \ \int_{(n-1)a}^{na} \d L \ \! P_D(L)\, .
\label{A.51.a}
\end{eqnarray}
This yields the PDF in the form
\begin{eqnarray} \label{disproportionation distribution}
P_D(L) \ &=& \
 \frac{w_T}{a} \! \ \exp\left(-\frac{w_T}{a} L\right)\, .
\end{eqnarray}

In the case of termination by combination, a chain of $n$ monomers arises from the
combination of two growing chains with lengths $m$ and $n-m$ ($1 \leq m \leq n-1$).
The distribution of chain lengths is then given by
\begin{equation}
p_C(n) \ =  \ \sum_{m=1}^{n-1} p(m) p(n-m) = w_T^2 (n-1) w_P^{n-2}\, ,
\end{equation}
and the corresponding density function is given in analogy with (\ref{A.51.a}) as
\begin{eqnarray} \label{combination distribution}
P_C(L) \ &=&
L \! \ \frac{w_T^2}{a^2} \! \ \exp\left(-\frac{w_T}{a} L\right)\, .
\end{eqnarray}

By comparing (\ref{disproportionation distribution}) and (\ref{combination distribution})
with the Gamma PDF (\ref{gamma_PDF}), we can identify
\begin{eqnarray}
P_D(L) \ = \
f_{{w_T}\!\mbox{\tiny{/}}{a},1}(L)\, ,
\end{eqnarray}
and
\begin{eqnarray}
P_C(L) \ = \
f_{{w_T}\!\mbox{\tiny{/}}{a},2}(L)\, .
\end{eqnarray}

The THC distribution parameters $q$ and $\beta(0)$ of the end-to-end vector distribution
(\ref{end-to-end vector distribution}) can be thus set as follows: termination by disproportionation
corresponds to $q_D=2$ and
\begin{eqnarray}
\beta_D(0) \ = \
\frac{a}{w_T}\, ,
\end{eqnarray}
while termination by combination to $q_C={3}/{2}$ and
\begin{eqnarray}
\beta_{C}(0) \ = \
\frac{2 a}{w_T}\, .
\end{eqnarray}

Combination and disproportionation are two competitive processes and do not occur to
the same extend for all polymers~\cite{Kamide,Hiemenz}.

\section*{Appendix B~\label{ap2}}

The uses of the path-integral representations (\ref{3.2}) and (\ref{3.7})
are not restricted only to the calculation of probability densities or partition functions.
These formulas can be also directly used to calculate the one-loop effective action,
i.e., Gibbs free energy, in QFT.
The one-loop effective action $\Gamma^{(1)}$ is defined as~\cite{IZ,Parisi,schubert}
\begin{eqnarray}
\Gamma^{(1)} \ = \ - \frac{1}{2} \left.\ln \det \left(
\frac{\delta^2S_e}{\delta \phi_i \ \delta\phi_j }
\right)\right|_{\phi_k = \phi_{k,c}} \, .
\end{eqnarray}
Here $S_e$ denotes the Euclidean field-theory action, $\{\phi_k \}$ represents the
constituent multiplet of scalar fields and $\phi_{k,c}$ is a solution of the {\em classical}
field equations of motion. For instance, for a relativistic complex massive scalar field that is minimally coupled to a background
electromagnetic field and self-interacts via potential $U(\phi_k)$ the Euclidean Lagrange density reads~\cite{IZ}
\begin{widetext}
\begin{eqnarray}
{\mathcal{L}}_e \ = \  \frac{1}{2}(\phi^*, \phi) \ \! \left[
                       \begin{array}{cc}
                         (\hat{p}_{\mu} + eA_{\mu})^2 + m^2 + U_{\phi^*\phi}(\phi_{k,c}) & U_{\phi^*\phi^*}(\phi_{k,c})\\
                         U_{\phi\phi}(\phi_{k,c}) & (\hat{p}_{\mu} - eA_{\mu})^2 + m^2 + U_{\phi\phi^*}(\phi_{k,c})  \\
                       \end{array}
                     \right]\ \!\left(
                                                                 \begin{array}{l}
                                                                   \phi\\
                                                                   \phi^*\\
                                                                 \end{array}
                                                               \right)
\end{eqnarray}
\end{widetext}
%
with $\hat{p}_{\mu} =  - i\partial_{\mu}$ and $U_{\phi^*\phi} =  \partial^2 U/ \partial \phi^* \partial \phi = U_{\phi\phi^*}$, etc. All scalar products are understood with respect to the Euclidean metric $\delta_{\mu\nu}$.
With the help of  the {\em Schur complement} technique for calculation of
determinants of partitioned matrices~\cite{Was} the ensuing one-loop contribution to the Gibbs free energy reads
\begin{eqnarray}
&&\mbox{\hspace{-7mm}}\Gamma^{(1)}[A] \ = \ - \frac{1}{2}\ln \det \left[ (\hat{p} + e A)^2 +
m^2 + U_{\phi^*\phi}\right]\nonumber \\[1mm]
&&\mbox{\hspace{-10mm}}- \frac{1}{2} \ln \det \left[(\hat{p} - e A)^2 +
m^2 + U_{\phi\phi^*} - U_{\phi\phi} {\mathcal{G}} \ \!U_{\phi^*\phi^*} \right] ,
\label{Ap.B.1a}
\end{eqnarray}
where ${\mathcal{G}} = \left[ (\hat{p} + e A)^2 +
m^2 + U_{\phi^*\phi}(\phi_{k,c})\right]^{-1}$ denotes the corresponding Green's function of the charged scalar particle in classical background fields $A_{\mu}$ and $U_{\phi^*\phi}(\phi_{k,c})$.

To illustrate  the connection with the THC statistics we consider for simplicity the situation with $U=0$. In this case we have
\begin{eqnarray}
\Gamma^{(1)}[A] = -\ \ln \det \left[ (\hat{p} + e A)^2 +
m^2 \right]\, .
\label{Ap.B.1b}
\end{eqnarray}
Note that the global factor $1/2$ has disappeared, because
\begin{eqnarray}
\det \left[ (\hat{p} - e A)^2 +
m^2 \right] &=& \det \left[ {\mathcal{C}}((\hat{p} - e A)^2 +
m^2){\mathcal{C}}^{-1} \right] \nonumber \\[1mm] &=& \det \left[ (\hat{p} + e A)^2 +
m^2 \right]\, ,
\end{eqnarray}
where the unitary operator ${\mathcal{C}}$ represents the charge conjugation operator, i.e.,  ${\mathcal{C}} \phi^{~}{\mathcal{C}}^{-1} = \phi^*$, ${\mathcal{C}}  \phi^* {\mathcal{C}} ^{-1} = \phi$ and ${\mathcal{C}} A^{\mu} {\mathcal{C}}^{-1} = - A^{\mu}$.

To calculate the functional determinant in (\ref{Ap.B.1b}) we utilize the method of the so-called $\zeta$-function regularization~\cite{Hawking}. The strategy is as follows;
%
we denote $\hat{\mathcal{A}} \equiv (\hat{p} + e A)^2 +
m^2$  and define the (spectral) $\zeta$-function $\zeta_{\mathcal{A}}(s)$ through the spectrum
$\{\lambda_n\}$ of $\hat{\mathcal{A}}$ as
\begin{eqnarray}
\zeta_{\mathcal{A}}(s) \ = \ \sum_{n} \frac{1}{\lambda_n^s}\, .
\label{Ap.B.2a}
\end{eqnarray}
With this the determinant in (\ref{Ap.B.1a}) can be calculated as
\begin{eqnarray}
\det \hat{\mathcal{A}} \ = \ \lim_{s\rightarrow 0}e^{-\zeta'_{\mathcal{A}}(s)}\, .
\label{Ap.B.3a}
\end{eqnarray}
%
Since the spectrum is typically not known,
we can instead of the defining relation (\ref{Ap.B.2a}) utilize the {\em heat kernel} method to
to compute $\zeta_{\mathcal{A}}(s)$.
The heat kernel ${\mathcal{G}}_{\mathcal{A}}(x,y,\beta)$ of the operator
$\hat{\mathcal{A}}$ is the fundamental solution of the heat-transfer type
equation
\begin{eqnarray}
\hat{\mathcal{A}}_x {\mathcal{G}}_{\mathcal{A}}(x,y,\beta)  \ = \ -
\frac{\partial}{\partial \beta} \ \! {\mathcal{G}}_{\mathcal{A}}(x,y,t)\, ,
\end{eqnarray}
with the Cauchy condition ${\mathcal{G}}_{\mathcal{A}}(x,y,\beta=0) =
\delta(x-y)$.

By writing the heat kernel  in  Dirac's bra-cket notation
\begin{eqnarray}
{\mathcal{G}}_{\mathcal{A}}(x,y,\beta) \ = \ \theta(\beta) \langle x |\exp(-\beta \hat{\mathcal{A}})|y \rangle\, ,
\end{eqnarray}
and using the identity
\begin{eqnarray}
\frac{1}{\lambda_i^s} \ = \ \frac{1}{\Gamma(s)} \int_0^{\infty} \d \beta \ \! \beta^{s-1} e^{-\beta \lambda_i}\, ,
\end{eqnarray}
valid for $\mbox{Re}(s) >0$,  we  see that
$\zeta_{\mathcal{A}}(s)$ can be alternatively written as the
Mellin transform of the trace of the heat kernel
${\mathcal{G}}_{\mathcal{A}}(x,y,\beta)$, namely
\begin{eqnarray}
\zeta_{\mathcal{A}}(s) \ = \  \frac{1}{\Gamma(s)} \int_{0}^{\infty}
\d \beta \ \beta^{s-1} \ \int_{-\infty}^{\infty} \d x \ {\mathcal{G}}_{\mathcal{A}}(x,x,\beta)\, .
\end{eqnarray}
%
%
Here the parameter $\beta$ is known as Schwinger'a proper time parameter.
By employing (\ref{Ap.B.3a}) we obtain
%
\begin{eqnarray}
\mbox{\hspace{-1mm}}\ln\det{\hat{\mathcal{{A}}}}  =  -\lim_{s\rightarrow 0} \frac{\d}{\d s} \!\!
\int_{0}^{\infty}\frac{\d \beta \! \ \beta^{s-1}}{\Gamma(s)}
 \int_{-\infty}^{\infty}\!\!\! \d x \ \!
{\mathcal{G}}_{\mathcal{A}}(x,x,\beta) ,
\label{eq2.3}
\end{eqnarray}
%
or equivalently [cf. (\ref{Ap.B.1a})]
\begin{eqnarray}
&&\mbox{\hspace{-8mm}}\Gamma^{(1)}[A] \ = \ \ln\det{\hat{\mathcal{{A}}}/M^2} \nonumber \\[2mm]
&&\mbox{\hspace{-8mm}}= \ \lim_{s \rightarrow 0}  \frac{\d}{\d s}
\int^{\infty}_{0}\! \frac{\d \beta  \! \ \beta^{s-1}}{\Gamma(s)} 
{\mbox{Tr}}\left[e^{-\beta((\hat{p}+ eA)^2 + m^2)/M^2}     \right]\! .
\end{eqnarray}
Here we have introduced a factor $M$ with the dimension of mass to maintain the argument of $\ln(\ldots)$ dimensionless.  In this case also $\beta$ is dimensionless.

Path integral enters when the functional trace ${\mbox{Tr}}(\cdots)$ is rewritten in the position-space representation, i.e.
\begin{eqnarray}
{\mbox{Tr}}(\cdots) \ = \ \int_{-\infty}^{\infty} \d x \ \! \langle x| \cdots |x \rangle\, ,
\end{eqnarray}
and when the PI representation
\begin{eqnarray}
&&\mbox{\hspace{-10mm}}\langle x | e^{-\beta(\hat{p} + eA)^2} | x \rangle  \nonumber \\[0mm]
&&\mbox{\hspace{-10mm}}= \ \int_{q(0) = q(\beta) =
x} {\mathcal{D}}q
 \exp\left[- \int_0^\beta \d \tau \ \! (\dot{q}^2/4 + i e\dot{q}A )    \right]\! ,
\end{eqnarray}
is utilized. With this we can finally write
%
%
\begin{eqnarray}
&&\mbox{\hspace{-9mm}}\Gamma^{(1)}\nonumber[A] \ = \  \lim_{s \rightarrow 0}
\frac{\d}{\d s}\ \frac{1}{\Gamma(s)} \int^{\infty}_{0} \frac{\d
\beta}{\beta} \ (M^2\beta)^s  e^{-\beta m^2} \nonumber \\[1mm]
&&\mbox{\hspace{7mm}}\times \ \! \oint {\mathcal{D}} q \ \exp\left[- S_e[{q},\dot{q}]\right]\, ,
\label{B.88}
\end{eqnarray}\\
%
where $S_e =  \int_0^{\beta}
\d \tau \
(\dot{q}^2/4 + i e\dot{q}A({q}) )$ is the corresponding quantum-mechanical Euclidean action.  Again,  the scalar products $\dot{q}_{\mu}\dot{q}^{\mu}$ and $\dot{q}_{\mu}A^{\mu}({q})$ are with respect to the Euclidean metric $\delta_{\mu\nu}$. 

Since the $\beta$-integral is generally not absolutely convergent, one cannot naively interchange derivation and limitation with integration.
So, strictly speaking, one must first evaluate the integral with the regulator $s>0$ and only at the end perform the differentiation in $s\rightarrow 0$. Fortunately, for many operators $\hat{\mathcal{A}}$ one can
uniquely evaluate this limit and hence analytically continue the path integral with $s>0$ to the path integral with the would-be $1/\beta$ term [formally ${\d}(M^{2s}\beta^{s-1}/{\Gamma(s)} )/\d s = 1/\beta + {\mathcal{O}}(s)$]. In fact, the latter path integral typically leads the so-called Schwinger determinant which is unregularized (infinite) expression and it is necessary to provide some  regulation scheme to obtain a well defined result.

The form (\ref{B.88}), in turn, allows to pinpoint yet another interesting r\^{o}le of the THC parameter,
namely $1/\varepsilon -1$  can be identified with a regulator $s$ in the $\zeta$-function regularization which, as we have just seen, is used in regularization of functional determinants [such as (\ref{Ap.B.1a})] in QFT.

In principle, one can proceed with the outlined THC statistics even to higher-loop orders in QFT calculations of Gibbs free energy by using the so-called world-line path integral representation. We shall not dwell into this approach here. The interested reader may consult, e.g., Ref.~~\cite{schubert}

%
%

\section*{References}
%

\end{document}